
\magnification 1200
\hoffset 1.0 truein
\hsize 6.2 truein
\voffset 1.0 truein
\vsize 8.7 truein

\def\doublespace{\baselineskip 24pt \lineskip 10pt \parskip 5pt plus 7 pt}
\def\today{\number\day\enspace
     \ifcase\month\or January\or February\or March\or April\or May\or
     June\or July\or August\or September\or October\or
     November\or December\fi \enspace\number\year}
\def\clock{\count0=\time \divide\count0 by 60
    \count1=\count0 \multiply\count1 by -60 \advance\count1 by \time
    \number\count0:\ifnum\count1<10{0\number\count1}\else\number\count1\fi}
\footline={\hss\rm -- \folio\ -- \hss}
\def\jref#1 #2 #3 #4 {{\par\noindent \hangindent=3em \hangafter=1
      \advance \rightskip by 5em #1, {#2}, {#3}, #4.\par}}
\overfullrule=0pt
\font\eightpt=amr8

\def\srm{\eightpt}

\def\title{\bf\twelvept}
\raggedright
\def\cmsq{\ifmmode{{\rm cm^{-2}}}\else{$\rm cm^{-2}$}\fi}
\def\cm3{\ifmmode{{\rm cm^{-3}}}\else{$\rm cm^{-3}$}\fi}
\def\kms{\ifmmode{{\rm km s^{-1}}}\else{$\rm km s^{-1}$}\fi}
\def\H2{\ifmmode{{\rm H_2}}\else{$\rm H_2$}\fi}
\def\HI{\ifmmode{{\rm H\>{\srm I}}}\else{{\rm H$\>${\srm I}}}\fi}
\def\sec{\ifmmode{^{\prime\prime}}\else{$^{\prime\prime}$}\fi}
\def\min{\ifmmode{^{\prime}}\else{$^{\prime}$}\fi}
\def\deg{\ifmmode{^{\circ}}\else{$^{\circ}$}\fi}
\def\point{\ifmmode{.\mskip-5mu}\else{$.\mskip-5mu$}\fi}
\def\solar{\ifmmode{_{\odot}}\else{$_{\odot}$}\fi}

\def\Msun{\ifmmode{M\solar}\else{{\it M\solar}}\fi}
\def\Lsun{\ifmmode{L\solar}\else{{\it L\solar}}\fi}

{\centerline {\bf DRAFT}}
\vskip 0.1 in
\centerline {\bf MOMENTUM TRANSFER BY ASTROPHYSICAL JETS}
\vskip 0.1 in
\centerline {L{\srm AWRENCE} C{\srm HERNIN} {\srm AND} C{\srm OLIN}
M{\srm ASSON}}
\centerline {Center for Astrophysics, MS 78, 60 Garden St, Cambridge, MA,
02178, USA}
\vskip 0.1 in
\centerline {E{\srm LISABETE} M. de G{\srm OUVEIA} D{\srm AL} P{\srm INO}}
\centerline {University of Sao Paulo, Instistuto Astronomico e Geofisico}
\centerline {Av. Miguel Stefano, 4200, Sao Paulo, SP 04301-904, Brazil}
\vskip 0.1 in
\centerline {\srm AND}
\vskip 0.1 in
\centerline {W{\srm ILLY} B{\srm ENZ}}
\centerline {University of Arizona, Steward Obs. and Lunar and Plan. Lab.,}
\centerline {Tuscon, AZ 85721, USA}
\vskip 0.2 in
\centerline {\today}
\vskip 0.1 in
\centerline {\bf {ABSTRACT}}
\vskip 0.1 in
\doublespace

We have used three dimensional smoothed particle hydrodynamical simulations to
study the basic properties of the outflow that is created by a protostellar
jet in a dense molecular cloud.  The dynamics of the jet/cloud interaction
is strongly affected by the cooling in the shocked gas behind the bow shock
at the head of the jet. We show that this cooling is very rapid, with the
cooling distance of the gas much less than the jet radius. Thus, although
ambient gas is initially driven away from the jet axis by the high thermal
pressure odf the post-shock gas, rapid cooling reduces the pressure and
the outflow subsequently evolves in a momentum-conserving snowplow fashion.
The velocity of the ambient gas is high in the vicinity of the jet head, but
decreases rapidly as more material is swept up. Thus, this type
of outflow produces extremely high velocity clumps of post shock gas which
resemble the features seen in outflows. We have investigated the transfer
of momentum from the jet to the ambient medium as a function of the jet
parameters. We show that a low Mach number (<6) jet slows down rapidly
because it entrains ambient material along its sides. On the other hand,
the beam of a high Mach number jet is separated from the ambient gas by a
low density cocoon of post-shock gas, and this jet transfers momentum
to the ambient medium principally at the bow-shock. In high Mach number jets,
as those from young stellar objects, the dominant interaction is therefore
at the bow shock at the head of the jet.
\vskip 0.2 in
\noindent {\it Subject Headings:} Hydrodynamics - stars: pre-main-sequence -
stars: mass loss - ISM: jets and outflows

\vfill\eject
\end